# Fluctuation spatial expansion and observational redshifts

Dale R. Koehler
82 Kiva Place
Sandia Park, New Mexico 87047

## Abstract

Classical determinations of galaxy distances and galaxy recessional velocities have been generated from luminosity and emission spectrometric data. The analyses of these galactic spectrometric electromagnetic frequency shifts have resulted in the Hubble law and are understood as a Doppler effect stemming from an expansion of space. In the present work, a galaxy-core expansion model with a time evolving matter and radiation distribution is put forth, leading to a supplementary treatment of the optical redshift measurements. Einstein's gravitational equations are assumed to apply within a galaxy-core spatial domain and, with utilization of a generalized Robertson-Walker-Schwarzschild metric, are used in order to calculate the evolutionary expansion characteristics. The galaxy-core model is described as a flat metric, matter plus radiation, $\sigma = 1/3$, energy distribution. It predicts an early time density fluctuation collapse, understood to be the formation of a galaxy hole, and provides an interpretational basis for the experimental data.

PACS: 04.70.Bw, 98.80.Hw

# 1.    INTRODUCTION

In the early part of the century, Hubble and Humason[1] cataloged the recessional velocities, relative to our Milky Way galaxy, of numerous galaxies, thereby identifying the velocity versus distance relationship now known as the Hubble law. Recessional velocities were calculable from analyses of the spectroscopic data associated with the individual light sources while distances were determined from the apparent magnitudes of these sources. It is now believed that these optical Doppler frequency shifts arise from a cosmological expansion of the intervening space itself[2], rather than from a motion of the galaxies through space[3,4].

In the present work, it is proposed that the observational expansion measurements derive from conditions in the local galactic environments. That is, the recessional velocity or redshift data are primarily measurements of the state of expansion in the emitting and observing galaxies. To further explore this description, the inter- and intra-galactic radiation emission, transmission, and detection processes are approximated into three regions of differing gravitational characteristics. For the emitting and observing galaxy regions, a galaxy-core model is developed to facilitate calculation of the pertinent expansion states and therefore the redshifts.

# 2.    GALAXY-CORE MODELING

Figure 1 illustrates a diagrammatic space and time pictorial of the galaxy, space, and radiation propagation conditions. Within the galactic regions, that is the emission and detection regions, an analysis of the propagation process requires examining the local expansion states since they effect the radiation. To this end, the galaxy-core model begins with an Einsteinian gravitational treatment of the early matter and radiation density fluctuation defining the core.

Einstein's gravitational equations[5], when qualified to an isotropic homogeneous space, have been solved by Friedmann[5] to yield the classic expanding universe equations for a perfect fluid, or continuous matter distribution;

$$G_{ab}(g_{ab}) \equiv R_{ab} - \frac{1}{2} g_{ab} R = 8\pi\, T_{ab} = 8\pi\big[\rho\, u_a u_b + p(g_{ab} + u_a u_b)\big] \quad . \quad (1)$$

$G_{ab}$ is the Einstein tensor, a function of the metric $g_{ab}$ and its first two derivatives, $R_{ab}$ is the Ricci tensor, $R$ the Ricci scalar, and $T_{ab}$ the energy-momentum tensor describing the material contents of the environment; $\rho =$ matter energy density and $p = constant\ x\ \rho = \sigma\rho =$ pressure energy density. The commonly used Robertson-Walker metric, for the three curved spaces is

$$ds^2 = -dt^2 + a(t)^2\left(1 + \frac{kr^2}{4}\right)^{-2}\Big[\,dr^2 + r^2 d\Omega^2\,\Big],\ where$$
$$k = +1\,,\ 0\,,\ -1\,, \tag{2}$$

for a sphere, plane or pseudosphere. The Friedmann solutions to these equations are

$$\frac{\ddot{a}}{a} = -\frac{4\pi}{3}(\rho + 3p) \quad and \quad \left(\frac{\dot{a}}{a}\right)^2 = \frac{8\pi}{3}\rho - \frac{k}{a^2} \quad , \tag{3}$$

where $a$ is the time evolving spatial expansion parameter and $k$ is the space curvature parameter.

As a similar model for the present calculations, the galaxy-core has been described in evolutionary terms as a continuous matter and radiation density fluctuation with $\sigma = 1/3$, with a flat space metric (producing an associated radius dependent attraction), and beginning at time $t_{core-birth}$. This is an evolving spatial and matter region, which subsequently begins collapsing (hole augmentation) at a time $t_{collapse}$. However, describing this spherically symmetric, radial coordinate centered, density fluctuation region requires a modified metric, which in the present treatment utilizes a combination of the Robertson-Walker form with a Schwarzschild form to yield

$$ds^2 = -\alpha(r) * dt^2 + a(t)^2 \left(1 + \frac{kr^2}{4}\right)^{-2} \left[\frac{dr^2}{\alpha(r)} + r^2 d\Omega^2\right] \quad . \tag{4}$$

The time dependence of the metric is contained exclusively in the expansion parameter $a(t)$ while $\alpha$ is a function only of the radial coordinate $r$. The system of equations represented by equation (1) is shown more explicitly in equation (5);

$$\begin{pmatrix} 1 & -g_{00}g^{11} & -g_{00}g^{22} & -g_{00}g^{33} \\ -g_{11}g^{00} & 1 & -g_{11}g^{22} & -g_{11}g^{33} \\ -g_{22}g^{00} & -g_{22}g^{11} & 1 & -g_{22}g^{33} \\ -g_{33}g^{00} & -g_{33}g^{11} & -g_{33}g^{22} & 1 \end{pmatrix}\begin{pmatrix} R_{00} \\ R_{11} \\ R_{22} \\ R_{33} \end{pmatrix} = -16\pi \begin{pmatrix} T_{00} \\ T_{11} \\ T_{22} \\ T_{33} \end{pmatrix} . \quad (5)$$

Subsequent calculation of the tensor elements $R_{ab}$ leads to equation set (6):

$$R_{00} = 3\frac{\ddot{a}}{a} - \frac{\alpha}{2a^2}\left[\frac{\alpha''}{\varepsilon} + \frac{\alpha'\varepsilon'}{2\varepsilon^2} + \frac{2\alpha'}{\varepsilon r}\right] = -4\pi\left[\rho - 3g_{00}p\right],$$

$$R_{11} = -\frac{\varepsilon}{\alpha^2}\left[a\ddot{a} + 2\dot{a}^2\right] + \frac{\alpha''}{2\alpha} + \frac{\alpha'}{\alpha}\frac{1}{r} + \frac{\varepsilon''}{\varepsilon} - \frac{\varepsilon'}{\varepsilon}\left[\frac{\varepsilon'}{\varepsilon} - \frac{1}{r} - \frac{\alpha'}{4\alpha}\right] =$$
$$\qquad -4\pi\,g_{11}\left[-g^{00}\rho - p\right],$$

$$R_{22} = -\frac{\varepsilon r^2}{\alpha}\left[a\ddot{a} + 2\dot{a}^2\right] + \frac{\alpha}{2}\left[\frac{\varepsilon''}{\varepsilon}r^2 + \frac{\varepsilon'}{\varepsilon}(3r + r^2\frac{\alpha'}{\alpha} - \frac{\varepsilon'}{\varepsilon}\frac{r^2}{2}) + 2(1 + r\frac{\alpha'}{\alpha})\right] - 1 =$$
$$\qquad -4\pi\,g_{22}\left[-g^{00}\rho - p\right],$$

$$R_{33} = -\frac{\varepsilon r^2 \sin^2(\theta)}{\alpha}\left[a\ddot{a} + 2\dot{a}^2\right] +$$
$$\qquad + \sin^2(\theta)\frac{\alpha}{2}\left[\frac{\varepsilon''}{\varepsilon}r^2 + \frac{\varepsilon'}{\varepsilon}(3r + r^2\frac{\alpha'}{\alpha} - \frac{\varepsilon'}{\varepsilon}\frac{r^2}{2}) + 2(1 + r\frac{\alpha'}{\alpha})\right] - \sin^2(\theta) =$$
$$\qquad -4\pi g_{33}\left[-g^{00}\rho - p\right],$$

$$with \quad g_{00} = \frac{1}{g^{00}} = -\alpha , \; g_{11} = \frac{a^2\varepsilon}{\alpha}, \; g_{22} = a^2\varepsilon r^2 \; and \; g_{33} = a^2\varepsilon r^2\sin^2(\theta) . \quad (6)$$

$R_{22}$ and $R_{33}$ are equivalent equations, reflecting the isotropic character of the metric of equation (4), while combination of $R_{11}$ and $R_{22}$ requires the restricting relationship between the variables $\alpha$ and $r$ displayed in equation (7);

$$\frac{\alpha''}{2\alpha} - \frac{1}{4}\frac{\alpha'}{\alpha}\frac{\varepsilon'}{\varepsilon} + \frac{1}{r^2}\left(\frac{1}{\alpha} - 1\right) = 0 \text{ , where } \varepsilon = \left(1 + \frac{kr^2}{4}\right)^{-2} . \qquad (7)$$

Primes denote differentiation with respect to the radial coordinate $r$. More explicitly, the three defining equations for $\alpha$ are

$$k = 0 \; ; \quad \alpha'' = \frac{2(\alpha-1)}{r^2} \quad ,$$

$$k = 1 \; ; \quad \alpha'' = \frac{2(\alpha-1)}{r^2} - \frac{\alpha'}{2}\frac{r}{1+\frac{r^2}{4}} \quad and$$

$$k = -1 \; ; \quad \alpha'' = \frac{2(\alpha-1)}{r^2} + \frac{\alpha'}{2}\frac{r}{1-\frac{r^2}{4}} \quad . \qquad (8)$$

The resultant expansion parameter equation is developed by combination of the $R_{00}$ and $R_{11}$ expressions to yield,

$$\left(\frac{\dot{a}}{a}\right)^2 = \frac{8\pi}{3}\rho + \frac{\alpha}{a^2}\left[\frac{1}{3\varepsilon}\left(\frac{\alpha''}{2} + \frac{\alpha'}{r} + \frac{\alpha'}{4}\frac{\varepsilon'}{\varepsilon}\right) - k\right] \quad ,$$

or,

$$\left(\frac{\dot{a}}{a}\right)^2 = H_0^2\left[\frac{\Omega_{s0}a_{s0}^3}{a^3} + \frac{\Omega_{r0}a_{s0}^4}{a^4}\right] + \frac{\alpha}{a^2}\left[\frac{1}{3\varepsilon}\left(\frac{\alpha''}{2} + \frac{\alpha'}{r} + \frac{\alpha'}{4}\frac{\varepsilon'}{\varepsilon}\right) - k\right] \quad . \quad (9)$$

$H_0$ is the Hubble constant and $a_{s0}$ is the normalized, present day, expansion parameter value. The $\alpha/a^2$, or $k(r)/a^2$, term is designated the curvature energy. Expansion factor $a$ is coupled to the radial position $r$ through the position dependent curvature energy and the metric factor $\alpha$, thereby predicting that the space-expansion rates will be variable throughout the matter distribution.

The most general solution to equation (7) for a flat metric generates the dependence

$$\alpha = 1 + A_1 / r + B_1 r^2 \quad .$$

(10)

Similar solutions are found for the $k = 1$ and $k = -1$ cases. Applying the flat metric solution for $\alpha$ to the region exterior to the total mass M, with the boundary condition at infinity, requires $B_1 = 0$ and results in zero curvature energy, consistent with the starting use of the flat metric. For the static case, comparison with the Newtonian potential form leads, independently, to the Schwarzschild metric factor,

$$\alpha = 1 - \frac{2GM / c^2}{r} \equiv 1 - \frac{r_s}{r} \quad .$$

(11)

Interior to the mass distribution, solution (10) contains a classical Newtonian-like $r^2$ term and a $1/r$ term, which are appropriate and necessary to describe distributions containing both a distributed mass and a central core mass element such as a hole. Following the development of Bergmann[6], that is, comparison with the linear Newtonian field theory, the distribution coefficient $B_1$ is determined to be $B_1 = (r_s - r_{hole}) / (2r_g^3)$. We require, as a boundary condition on the metric factor $\alpha$, equality across the mass radial boundary $r_g$ (the radius of the total core mass M). Then solution (10) yields equation (12) for $\alpha$ and equation (13) for the "curvature" inside the matter distribution,

$$\alpha = 1 - \left[\frac{3r_s - r_{hole}}{2r}\right] + \frac{r_s - r_{hole}}{2r_g^3}r^2 \, , \tag{12}$$

$$k = \alpha B_1 = \left[1 - \left[\frac{3r_s - r_{hole}}{2r}\right] + \frac{r_s - r_{hole}}{2r_g^3}r^2\right]\frac{r_s - r_{hole}}{2r_g^3} \, . \tag{13}$$

The Schwarzschild hole radius is denoted by $r_{hole}$. Therefore, although the flat space description holds at large $r$ exterior to the matter, the metric produces an attractive, positive-curvature-like behavior in the interior of the mass distribution.

For other than flat spaces and in the regions external to the mass, the resulting $k(r)$ values are $k(r) = -\alpha^2$ for $k = +1$ and $k(r) = \alpha^2$ for $k = -1$. The mass interior region could also be treated with either the $k = +1$, or $k = -1$ descriptions, however, the initial density conditions are assumed those of a flat space.

The time dependent expansion parameter, density and temperature solutions for equation (9) (curvature $k = -\varepsilon(r) = B_1\alpha(r)$, $\sigma = 1/3$) are,

$$t - t_{core-birth} = \frac{C_1}{\varepsilon^{3/2}}\left[\begin{array}{l}\sin^{-1}\left(\sqrt{\frac{2a\varepsilon - 1 + \kappa}{2\kappa}}\right) - \sqrt{\frac{\kappa^2 - 1}{4} + \frac{a\varepsilon}{C_1} - \left(\frac{a\varepsilon}{C_1}\right)^2} \\ + \sqrt{\frac{\kappa^2 - 1}{4}} - \sin^{-1}\left(\sqrt{\frac{\kappa - 1}{2\kappa}}\right)\end{array}\right];$$

$$\varepsilon = \varepsilon(r) \, ; \quad C_1 = \Omega_{s0}H_0^2 a_{s0}^3 = \frac{t_{collapse} - t_{core-birth}}{\pi/2 + \eta_0}\varepsilon(r_{hole})^{3/2} \, ;$$

$$\eta_0 = \sqrt{\frac{\kappa_{rhole}^2 - 1}{4}} - \sin^{-1}\left(\sqrt{\frac{\kappa_{rhole} - 1}{2\kappa_{rhole}}}\right) \; ; \; \Omega(t) = \Omega_{s0}\left(\frac{a_{s0}}{a}\right)^3 + \Omega_{r0}\left(\frac{a_{s0}}{a}\right)^4 \; ;$$

$$\kappa = \sqrt{\frac{4\Omega_r \varepsilon\, a_{s0}}{C_1} + 1} \; ; \quad Temp(t) = \frac{T_0}{a} = T_{experimental}(today)\frac{a_{core}(today)}{a} \; . \quad (14)$$

The zero subscripts for the density $\Omega_{s0}$ and the temperature $T_0$ (as for the Hubble constant $H_0$ and the expansion parameter $a_{s0}$) denote present day values. Radiation energy density ratio $\Omega_r = \Omega_{r0}/\Omega_{s0}$ expresses the present-day fractional radiation energy component (radiation energy/matter energy). Although the time value for the region's initial collapse, $t_{collapse}$, (with radial coordinate $r_{hole}$) is an input to the solution, absent a physical mechanism to explain an inhibition of collapse, the spatial region at the center of the mass, at $r = r_{origin} \equiv r_{hole}$, will begin to collapse immediately at $t = t_{core-birth}$. Under these circumstances, the qualitative expansion behavior is not affected but quantitatively the collapse process merely begins at the earlier time.

The approximate time at which collapse occurs, at a given mass radius, is given by the expression,

$$t_{collapse}(r_i) \cong t_{collapse}(r_{origin}) * (\text{ curvature }(r_{origin})/curvature(r_i))^{3/2} \; ,$$

where $r_i$ is the radius at the mass shell in question and $r_{origin}$ is the beginning spatial collapse zone radius at $t_{1st\, collapse}$. A growing hole results from the collapsing regions at radii less than the singularity surface and an initial seed-hole mass is determinable from the present-day experimentally measured hole-mass value through the collapse time expression and equation (15);

$$Hole\ mass(today) = (r_{collapse}(today)^3 - r_{collapse}(birth)^3) * Galaxy\ core\ mass * \left(\frac{1}{r_g}\right)^3$$
$$+ Seed\ Hole\ mass\ . \qquad (15)$$

Since the present model predicts collapse without the presence of a seed mass, we proceed without the assumption of an initial seed. The crossover point of the modified metric is the singular spherical surface at

$$r_{\sin gularity} = \left[\sqrt{\left(\frac{r_g}{3r_s/2}\right)^3 + \left(\frac{3}{2}\right)^2} + \frac{3}{2}\right]^{1/3} - \left[\sqrt{\left(\frac{r_g}{3r_s/2}\right)^3 + \left(\frac{3}{2}\right)^2} - \frac{3}{2}\right]^{1/3} \qquad (16)$$

inside of which $\alpha$ is negative and the "curvature term" is positive. A negatively curved, approximately flat, region exists between the singularity surface and $r_g$. The metric factor $\alpha$ is plotted in Fig. 2.

Bounds on the galaxy-core birth time are forthcoming from consideration of the galaxy-core mass radius and the hole mass extremum at the singularity radius. If the initial mass distribution's maximum radius at time $t_{core-birth}$ is less than or equal to the singularity radius for that mass, then all of the region will eventually collapse but the collapse process and the mass distribution itself will be unobservable to regions outside the Schwarzschild radius, which is the singularity radius in this case. Model cases with smaller masses or later galaxy birth times will additionally contain the second, more rapidly expanding, negatively curved region mentioned above. A mass distribution with a mass-radius equal to the singularity radius possesses the interesting characteristic that the curvature of the region at the mass-space boundary is zero, thereby exhibiting equality across the boundary of both the curvature factor and the metric factor $\alpha$.

If one assumes that present day galaxy holes derive from such a collapse process and that the present day hole mass (neglecting accretion processes) represents the current collapsed mass, then for the two region model, an upper bound to the birth-time can be calculated by utilizing the radius of the collapsed mass today. The calculation sets the collapsed mass radius, which equals $r_g *$ *(hole mass ratio)$^{1/3}$*, equal to the singularity radius. This upper time bound is calculated to be 0.32 years, for a galaxy-core mass of $10^9$ stellar masses (density $\Omega_{s0}$ relates the galaxy-core radius to the galaxy-core mass) using equation (17) for the space expansion factor for a "radiation plus matter" universe and 1.5% for the hole-mass/core-mass ratio. For mass and birth-time combinations greater than the upper extremum, the present day collapsed mass will be less than the experimental hole value referred to above. The calculational result for the lower time bound for the galaxy-core birth time (resulting for a galaxy-core mass radius that approaches the singularity radius) is 0.05 years. This dependence of birth times on galaxy-core mass is illustrated in Fig. 3. Figure 4 displays the growing collapse zone as a function of time for a galaxy-core birth time of 0.32 years or a present-day collapsed mass radius of $0.99 * 3/2 * r_s$. Figure 5 shows the range of collapse times as a function of the galactic-core radial coordinate. For birth times close to the upper extremum, the model exhibits a linear core-mass dependence (see Fig. 6) in good agreement with the experimental latter-day galaxy hole measurements; Kormendy[7], Magorrian[8], and Kormendy[9] have shown such a galaxy bulge-mass correlation (also see Fig. 2 in Gebhardt[10]).

The variable $T_{experimental}$ is the cosmic microwave background temperature and is utilized as the intra-galaxy reference. It reflects the anticipated impact of a different expansion factor for intragalaxy space and intergalaxy space. Temperature is assumed to follow an $a^{-1}$ behavior throughout; after collapse, however, since the expansion factor loses its definition, an arbitrary value is assigned which is derived from the first collapsing mass-shell radius, that is

$$a_{collapse} = a_{core\text{-}birth} \left( \Omega_{\text{Gal core}}(t_{core\text{-}birth}) / \Omega_{\text{hole}}(t_{1st\ collapse}) \right)^{1/3} .$$

The cosmological density, $\Omega_{s0}$, and other model parameters are either experimental or observationally estimated. For the following expansion parameter, density and temperature calculations, an early birth-time perspective is used where $t_{core\text{-}birth} = 0.32$ years, galaxy-core mass $= 10^9$ stellar masses, $\Omega_{s0} = 1$, $\rho_{critical} = 7.2 \times 10^{-27}$ kg m$^{-3}$ and Hubble's constant $H_0 = 0.485 \times 10^{-10}$/year producing an age of $13.7 \times 10^9$ years. In the spirit of the present model, the Hubble constant is interpreted as a local galactic-based expansion measurement rather than an intergalactic spatial measurement. Also in the present model, for a $t^{2/3}$ time behavior of the expansion parameter, $t(now) \approx 2/3 * H_0^{-1}$.

Before time $t_{core-birth}$ the universe is assumed governed by an expansion, space density and temperature behavior determined by radiation and matter energy;

$$\left(\frac{t}{age}\right) = \frac{1}{\sqrt{\Omega'_{s0}}}\left[\left(\frac{a'}{a'_0} - 2\frac{\Omega'_{r0}}{\Omega'_{s0}}\right)\left(\frac{a'}{a'_0} + \frac{\Omega'_{r0}}{\Omega'_{s0}}\right)^{.5} + 2\left(\frac{\Omega'_{r0}}{\Omega'_{s0}}\right)^{1.5}\right],$$

$$a'_0 \equiv a_{space}(today) = a_{s0} \ , \quad \Omega'(t) = \Omega'_s(a'_0)\left(\frac{a'_0}{a'(t)}\right)^3 + \Omega'_r(a'_0)\left(\frac{a'_0}{a'(t)}\right)^4, \quad and$$

$$Temp'(t) = T_0\left(\frac{a_{core}(today,r)}{a(t,r)}\frac{a(t,r)}{a'(t)}\right), \tag{17}$$

where $\Omega'_{s0} \equiv \Omega'_s(a'_0)$ and $\Omega'_{r0} \equiv \Omega'_r(a'_0)$. Radiation modifies the time evolution but the intergalaxy, or space, expansion is the approximate $t^{2/3}$ behavior of the $k = 0(zero)$ flat universe. Temperature reflects the impact of an observer measuring an intergalactic expansion-state parameter $a'(t)$ from within a galaxy with expansion parameter $a(t, r)$.

The presence at time $t_{core-birth}$ of a density fluctuation, begins the departure from relative uniformity which describes the early universe with $a_{core}(t_{core-birth}) = a_{intergalaxy}(t_{core-birth})$. In the present perfect fluid modeling, the description provided by equation (9) only incorporates the gravitational physics of the collapsing galaxy-core space. Any subsequent matter accretion processes, or other energy sources, are not included. The continuous matter and radiation energy of the early universe, constituting the galaxy-core material, is considered the perfect fluid used to determine the expansion factor solution form. Figures 7-9 display the results of the calculations for the evolutionary character of the galaxy-core expansion parameter, the galaxy-core density function, and the galaxy-core temperature. The marked behavior at $t_{collapse}(r)$ reflects the collapse of the space, and the associated matter and radiation energy, responsible for hole augmentation. A present hole mass value, as mentioned earlier, equal to 1.5% of the galaxy bulge-mass (we have associated the galaxy bulge-mass with our core-mass), has been used for the growing hole. Its radius is assumed to be the Schwarzschild value. A terminal radius after collapse other than the Schwarzschild value is possible but not predictable from the model. Moreover, the concept of a collapsed spatial singularity has not been otherwise treated in the present calculational analysis. For Figures 7-9, the hole-radius unit value is that of the hole at the time $t = 2\ t_{core-birth}$. Both a spatial "z" definition, $z = a_{spa}\ (today)/a_{spa}\ (t) - 1$, and a galaxy-observer "z" definition, $z = a_{obs}\ (today,\ r_{obs})/a_{obs}\ (t,\ r_{obs}) - 1$ are plotted in Fig. 8. The galaxy-observer is positioned at $0.75 \times$ core-radius and because the expansion parameter in the outer region of the model core-mass distribution behaves quite differently from the exterior spatial region, there is a significant difference in the redshift functions. All of the galaxy-core region internal to the singularity shell-radius is predicted to collapse eventually but the collapse time follows an inverse curvature dependence which results in infinitely long collapse times, $t_{collapse}(r) = \pi\, C_1\, /\, k(r)^{3/2}$, as the curvature, $k$, approaches zero at the singularity transition edge. The time evolution of the galaxy-core expansion parameter, in the outer regions of the mass distribution, follows the negative curvature behavior,

$$t - t_{core-birth} = \frac{C_1}{\varepsilon^{3/2}}\left[\begin{array}{c}\sqrt{\kappa + \dfrac{a\varepsilon}{C_1} + \left(\dfrac{a\varepsilon}{C_1}\right)^2} - \dfrac{1}{2}\ln\left(\sqrt{\kappa + \dfrac{a\varepsilon}{C_1} + \left(\dfrac{a\varepsilon}{C_1}\right)^2} + \dfrac{a\varepsilon}{C_1} + \dfrac{1}{2}\right) \\[10pt] -\sqrt{\kappa} + \dfrac{1}{2}\ln\left(\sqrt{\kappa} + \dfrac{1}{2}\right)\end{array}\right];$$

$$\varepsilon = \varepsilon(r)\;;\; C_1 = \Omega_{s0} H_0{}^2 a_{s0}{}^3\;;$$

$$\kappa = \sqrt{\frac{\Omega_r \varepsilon\, a_{s0}}{C_1}}\;\;;\;\;\; \Omega(t) = \Omega_{s0}\left(\frac{a_{s0}}{a}\right)^3 + \Omega_{r0}\left(\frac{a_{s0}}{a}\right)^4\;;$$

$$Temp(t) = \frac{T_0}{a} = T_{\exp erimental}(today)\frac{a_{core}(today)}{a}\;\;. \tag{18}$$

After a region collapses, the galaxy-core density function for that region is the hole density function and, as displayed in Fig. 8, the density is that of the growing Schwarzschild-radius hole.

## 3. OBSERVATIONAL REDSHIFT MODELING

For the radiation emission and detection process, the photon propagation behavior along the time evolving emitter to observer path, is determined by the equation for null geodesics and integration over time along the light path. This leads to the expression relating the emitted and observed time intervals[11],

$$\int_{t_e}^{t_{e2}} dt/a = \int_{t_o}^{t_{o2}} dt/a \; ; \; t_{e2} = t_e + \Delta t_e \; ; \; t_{02} = t_0 + \Delta t_0 \; . \tag{19}$$

The resulting Friedmann-Robertson-Walker (FRW) redshift expression[11], is

$$z = \frac{\nu_e - \nu_o}{\nu_o} = \Delta t_o / \Delta t_e - 1 = a(t_o)/a(t_e) - 1 \tag{20}$$

and illustrates that only the initial and final expansion states, $a_j(t_i)$ (i = emitter time or observer time, j = emitter or observer), determine the overall resultant change in frequency or wavelength of the propagating radiation. In other words, in a non-monolithic universe where local warping is present and where radiation emission sources and radiation observers (detectors) are both embedded in such locally warped regions, calculation of the radiation modification (redshift) involves calculation of the local region's expansion state as manifest in the local expansion parameter $a_j(t_i)$. The potential energy, or wavelength, diagram illustrated in Fig. 10 is a pictorial representation of the evolving photon energy state as it propagates (1), through the emitting galaxy, (2), out of the galaxy-space interface, (3), through the spatially expanding path between emitting and observing galaxies, (4), into the observer space-galaxy interface, and finally, (5), to the detection point within the observer galaxy. The observer galaxy and the emitter galaxy are assumed to exhibit the same time evolutionary or expansion characteristics. In such a path, the wavelength stretching (photon energy loss) step at the emitter galaxy-space interface and the energy loss process during intergalactic travel is mirrored at the second observer space-galaxy interface where the energy loss is partially recovered and the photon wavelength decreases. The expanding galaxy regions therefore produce the net overall energy change, or photon wavelength increase, during the time interval from emission to detection. If the $a_{emitter}$ and $a_{observer}$ evolution lines corresponded to the same $t^{2/3}$ time behavior as $a_{space}$, then no energy loss or recovery would be incurred at the space-galaxy interfaces. Although the present model calculations are limited to the core expansion-factor time-development, the notion of local space warping and its impact on the propagating radiation is still appropriate.

If the galaxy environment influences radiation redshifting, then microwave background radiation (CBR) is also affected and should display a wavelength offset equivalent to the ratio between present-day galaxy-space and intergalaxy-space, expansion parameter values;

$$CBR\ offset \equiv \left( \frac{a_{galaxy}(r,\,today)}{a_0'} \right) = \left( \frac{a_{galaxy}(r,\,today)}{a_{space}(today)} \right) \quad ;$$

$$Actual\ temp_{CBR} = Measured\ temp_{CBR} * CBR\ offset \quad . \tag{21}$$

A radiation evolution and propagation diagram is shown in Fig. 11 and, as illustrated, suggests an actual inter-galaxy background radiation temperature lower (longer wavelength) than measured inside the galaxy. However, since $a_{space}(today)$ is probably greater than that in the outer regions of the galaxy, a less-than-one CBR offset is predicted for observers in these regions. That is, appropriate galaxy models, with observers in high mass-density regions, would predict a smaller intergalaxy CBR temperature than that which is measured, a phenomenon that derives from the local space warping produced by the galaxy matter. For an observer in the outer regions (at $0.75 * r_g$) of the core-mass distribution used here, for example, the offset is 9.6.

## 4.    SUMMARY

In conclusion, the galaxy-cores, as modeled, display radius-dependent expansion rates and exhibit time evolution rates greater than $t^{2/3}$ dependencies in the outer regions while the inner regions are collapsing. We postulate from this modeling that cosmological redshift data should be interpreted as measurements of a localized galactic expansion parameter, both at the emitter and at the observer, and that cosmic microwave background radiation measurements should be impacted by the difference between galaxy and intergalaxy expansion rates.

## 5.  APPENDIX

As a second consideration of a galaxy-core birth-time minimum, a light crossing time interval, or gravitational propagation time interval, associated with coherent evolution of the mass fluctuation is required. Calculation of this equilibration time interval as $t_{equil} = 2r_g/c$ leads to a core-birth time approximately that of the upper-bound birth-time extremum. Positing that $t_{equil} \cong t_{upper\ extremum}$ then constrains the hole-mass/core-mass ratio (no longer a free model parameter). Therefore, with $m_{h\text{-}c} \equiv$ *hole-mass/core-mass ratio*,

$$t_{equil} = \frac{2r_g}{c} = \frac{2r_{g0}}{c}\frac{a^{'}}{a_0^{'}} = t_{upper\ extremum} \; ;$$

$$t_{upper\ extremum} = \frac{age}{\sqrt{\Omega^{'}_{s0}}}\left[\left(\frac{a^{'}}{a_0^{'}} - 2\frac{\Omega^{'}_{r0}}{\Omega^{'}_{s0}}\right)\left(\frac{a^{'}}{a_0^{'}} + \frac{\Omega^{'}_{r0}}{\Omega^{'}_{s0}}\right)^{.5} + 2\left(\frac{\Omega^{'}_{r0}}{\Omega^{'}_{s0}}\right)^{1.5}\right] \cong$$

$$\cong \frac{age}{\sqrt{\Omega^{'}_{s0}}}\left[\frac{3}{4}\frac{(a'/a_0^{'})^2}{\sqrt{\Omega^{'}_{r0}/\Omega^{'}_{s0}}}\right],$$

where the radiation terms dominate at the early times. We have defined the upper extremum such that the collapsed mass radius, which equals $r_g *$ *(hole mass ratio)$^{1/3}$*, is equal to the singularity radius and therefore, utilizing the curvature equation (13), we get

$$\frac{r_s}{r_g} = \frac{2GM}{c^2}\left(1 + \frac{\Omega'_{r0}}{\Omega'_{s0}}\frac{a_0'}{a'}\right)\left(\frac{a_0'}{a'}\right)\frac{1}{r_{g0}} = \frac{2m_{h-c}^{1/3}}{3 - m_{h-c}} \equiv M_{h-c} \, ,$$

$$\frac{a'}{a_0'} \cong \sqrt{\frac{2GM/c^2}{M_{h-c}r_{g0}}\frac{\Omega'_{r0}}{\Omega'_{s0}}} \, ,$$

and the resultant hole-mass/core-mass ratio expression,

$$m_{h-c} = \left(\frac{9}{16}\pi\rho_c G \, age^2\right)^3 \quad . \qquad (A1)$$

With such a gravitational coherence-time requirement and the merging of the lower-bound birth timeline into the upper-bound birth timeline, all mass distribution fluctuations in the model are then described as the two-region type, being born along an upper-bound extremum timeline (as in Fig. 3), which is approximately, in the early radiation-dominant era (< approx. $10^4$ years; see equation (17)),

$$t_{core-birth} \cong \frac{16}{3}\left(\frac{r_{g0}}{c}\right)^2\frac{\sqrt{\Omega'_r}}{age} \quad . \qquad (A2)$$

The calculational result for $m_{h-c}$ is $m_{h-c} = 3.9 \times 10^{-3}$. Since no accretion processes are included, the result should be compared (as a lower limit value) to the experimental data of Magorrian[8], where $\log(M_{hole}/M_{bulge}) = -2.28$ (mean with std.dev. = 0.51) or $M_{hole}/M_{bulge} = 5.2 \times 10^{-3}$.

## 7. FIGURE CAPTIONS

FIG. 1. Space-Time Diagram for galaxies in the evolving universe.

FIG. 2. Flat space metric factor $\alpha$ (absolute value) for a mass-region consisting of a geometrically centered mass distribution. The radial coordinates are in hole-radii units and are displayed at $t_{core\text{-}birth}$ ($r_g = 0.2$ and $r_{origin} = 1.4 \times 10^{-8}$ light-years).

FIG. 3. Extremum time lines, $t_{max}$ and $t_{min}$, for galaxy-core birth. Galaxy-core birth times are in years. Galaxy-core mass units are in stellar masses.

FIG. 4. Collapse zone mass ratio (hole mass ratio), $M_c$, versus time for a galaxy-core birth time of 0.32 years and a core-mass of $10^9$ stellar masses.

FIG. 5. Collapse times versus the galaxy-core radial coordinate (in units of hole-radii and displayed at galaxy-core birth time). The galaxy-core birth time is 0.32 years and the core mass is $10^9$ stellar masses.

FIG. 6. Hole-mass versus core-mass (bulge-mass). The experimental data in Gebhardt [10] are compared with model calculations employing a core-birth time of 0.32 years. Hole- and bulge-masses are in stellar mass units.

FIG. 7. Evolutionary Time (years) Development of the evolving expansion factor, $a_{time, r}$, from time $t_{core\text{-}birth}$ to the present. Radial coordinate values "b"(cross),

"c"(diamond), "d"(box) and "e"(plus symbol) correspond to 1, $10^4$, $2.2 \times 10^6$ and $10^7$ hole radii (hole radius = $1.4 \times 10^{-8}$ light-years). The "e" radius occurs at $0.75 \times$ galaxy core radius. The hole expansion factor, $a_{black\ hole}$, has been multiplied (increased) by a factor of $10^{40}$ for presentation in the figure.

FIG. 8. Evolutionary Time (years) Development of the relative density, $\Omega\_Gal_{time,\ r,}$ from time, $t_{core-birth}$ to the present. Radial coordinate values "b"(cross), "c"(diamond) and "e"(plus symbol) correspond to 1, $10^4$ and $10^7$ hole radii (hole radius = $1.4 \times 10^{-8}$ light-years). The "e" radius occurs at $0.75 \times$ galaxy core radius. Hole densities have been multiplied (reduced) by a factor of $10^{-100}$ for display purposes. The intergalaxy density, $\Omega\_spa$ (solid), z_space (solid) and z_observer (cross) are also shown.

FIG. 9. Evolutionary Time (years) Development of the temperature (K), $T_{time,\ r,}$ from time, $t_{core-birth}$ to the present. Radial coordinate values "b"(cross), "c"(diamond) and "e"(plus symbol) correspond to 1, $10^4$ and $10^7$ hole radii (hole radius = $1.4 \times 10^{-8}$ light-years). The "e" radius occurs at $0.75 \times$ galaxy core radius. The hole temperature is $7.5 \times 10^{52}$ K and has been multiplied (reduced) by a factor of $10^{-40}$ for display purposes. The intergalaxy temperature, $T\_spa$ (solid), is also shown.

FIG. 10. Potential Energy or Wavelength Diagram for photon propagation along a galaxy-emitter to galaxy-observer path.

FIG. 11. Potential Energy or Wavelength Diagram for photon propagation along a microwave-emitter to galaxy-observer path.

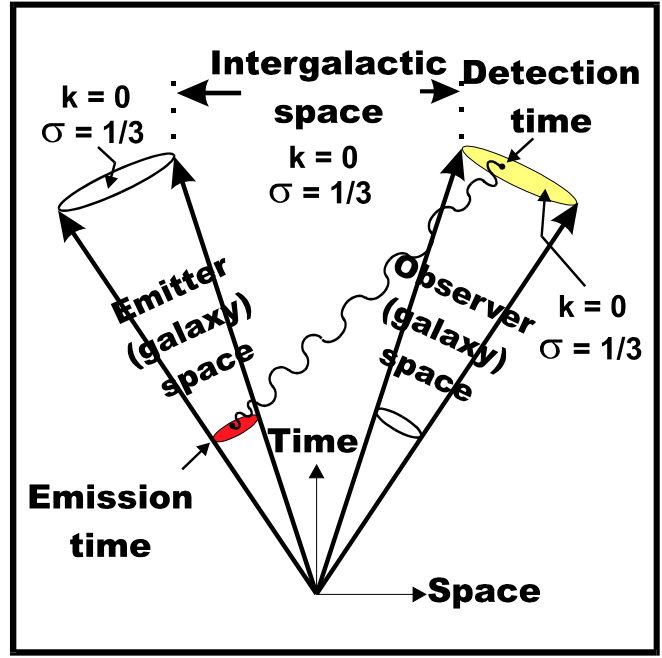

Figure 1  Dale R. Koehler  Fluctuation spatial expansion and observational
redshifts

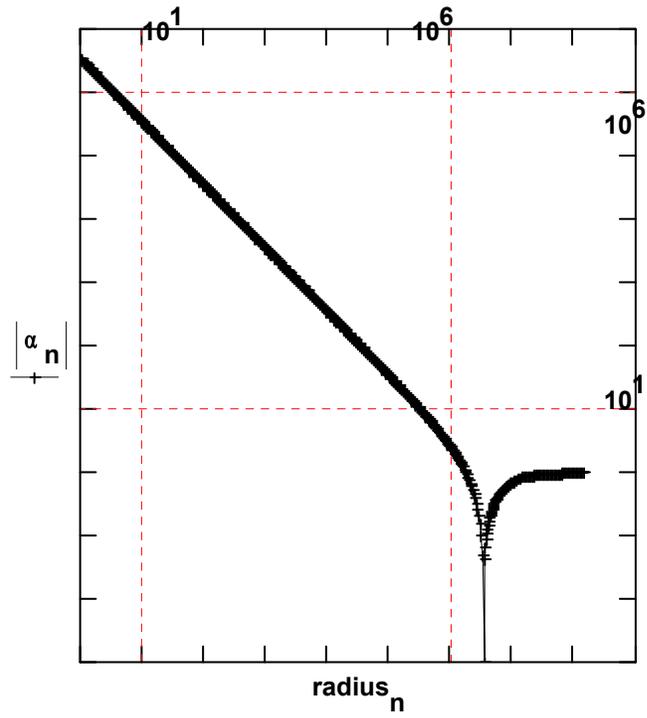

Figure 2  Dale R. Koehler  Fluctuation spatial expansion and observational redshifts

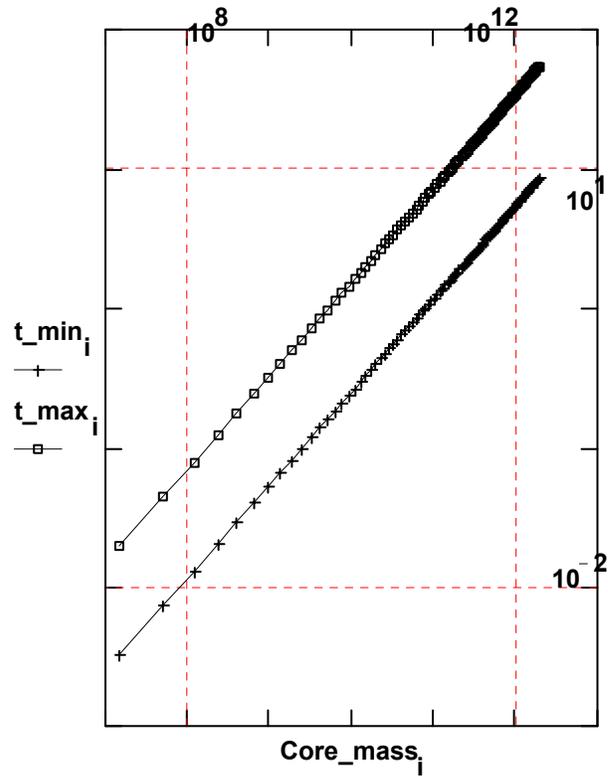

Figure 3  Dale R. Koehler  Fluctuation spatial expansion and observational redshifts

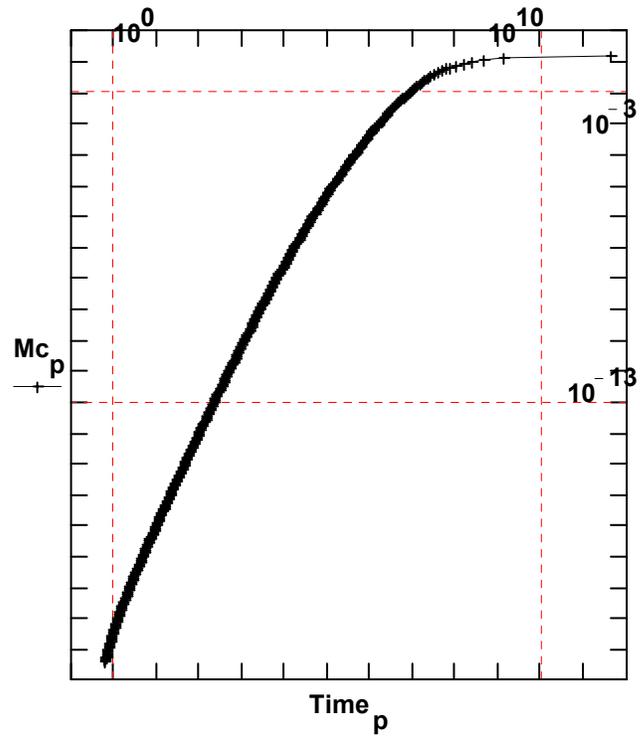

Figure 4  Dale R. Koehler  Fluctuation spatial expansion and observational redshifts

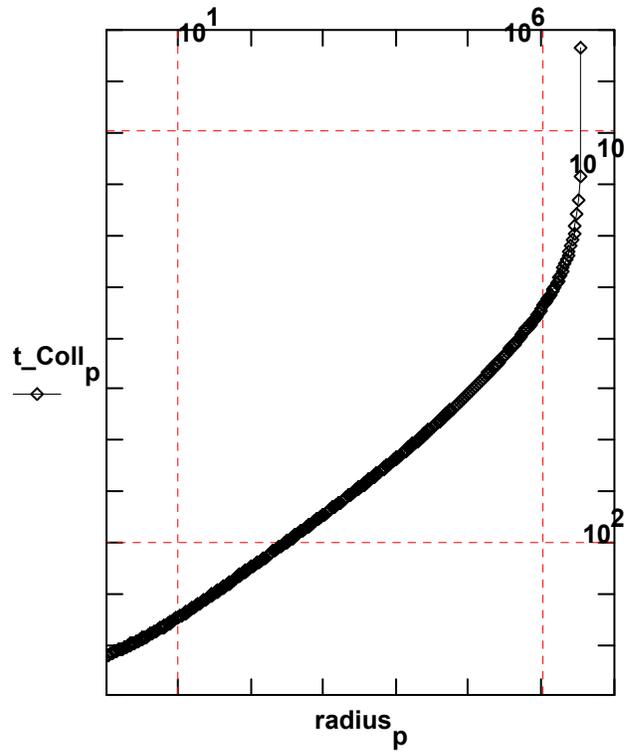

Figure 5  Dale R. Koehler  Fluctuation spatial expansion and observational redshifts

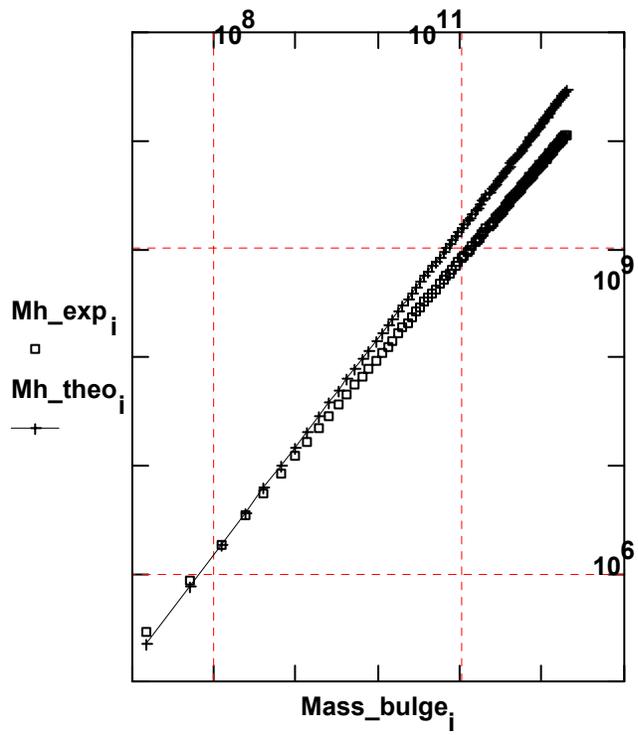

Figure 6  Dale R. Koehler  Fluctuation spatial expansion and observational redshifts

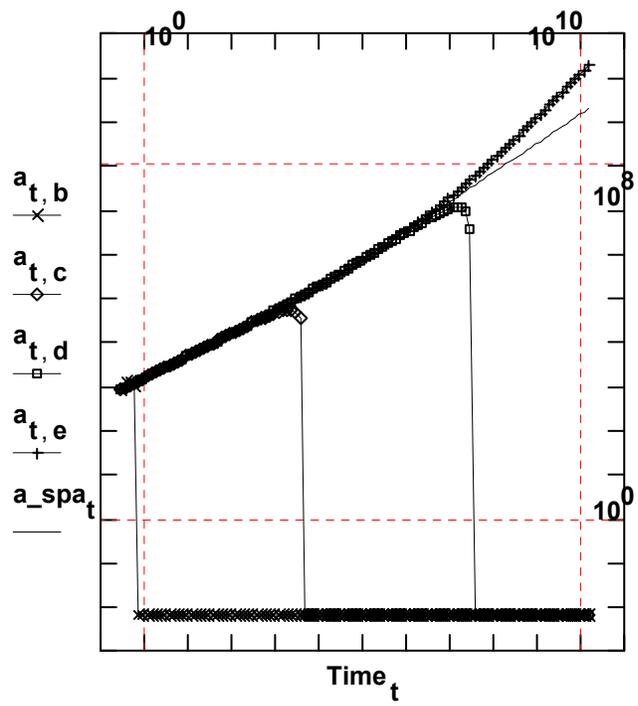

Figure 7  Dale R. Koehler  Fluctuation spatial expansion and observational redshifts

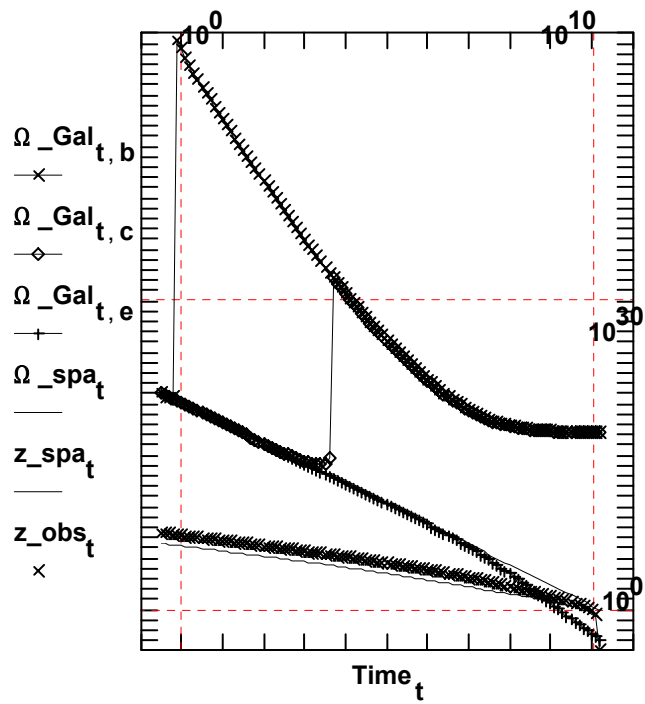

Figure 8  Dale R. Koehler  Fluctuation spatial expansion and observational redshifts

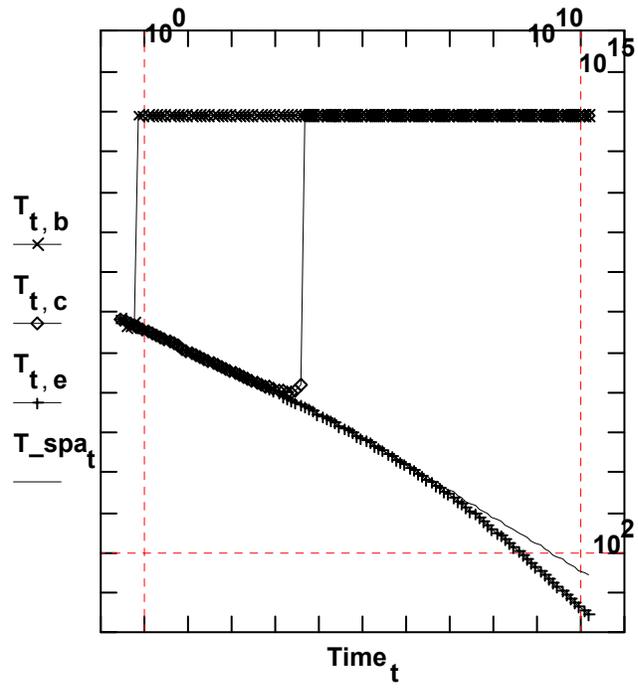

Figure 9  Dale R. Koehler  Fluctuation spatial expansion and observational redshifts

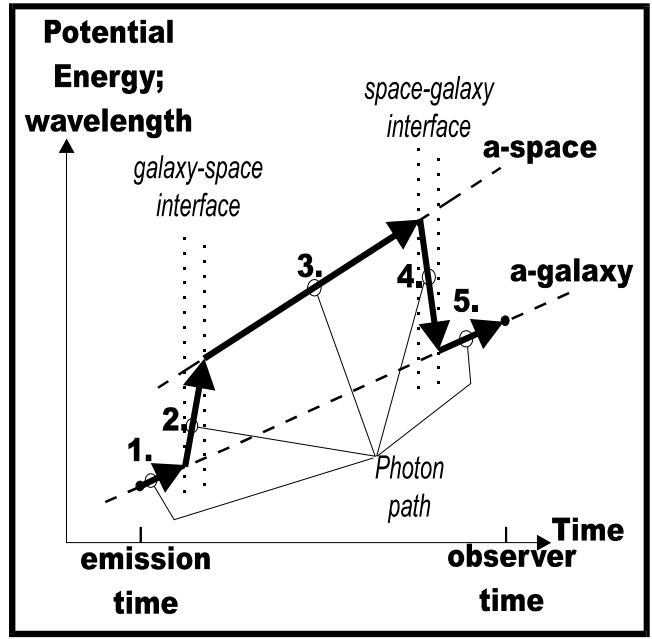

Figure 10  Dale R. Koehler  Fluctuation spatial expansion and

observational redshifts

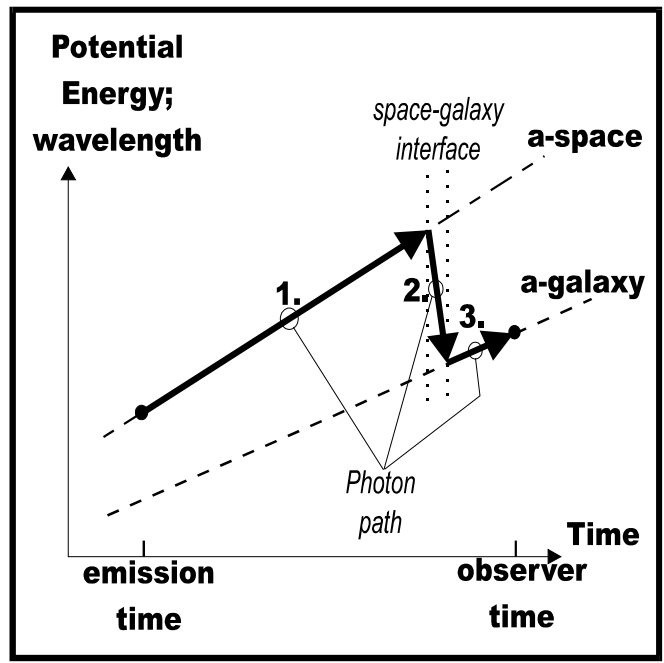

Figure 11  Dale R. Koehler  Fluctuation spatial expansion and observational redshifts